# Diode-pumped 88-fs SESAM mode-locked Tm,Ho:CLNGG laser at 2090 nm

ANNA SUZUKI,[1,*] YICHENG WANG,[1] SERGEI TOMILOV,[1] ZHONGBEN PAN[2], AND CLARA J. SARACENO[1]

[1]*Photonics and Ultrafast Laser Science, Ruhr Universität Bochum, Universitätsstraße 150, 44801 Bochum, Germany*
[2]*School of Information Science and Engineering, Key Laboratory of Laser and Infrared System of Ministry of Education, Shandong University, Qingdao 266237, China*
*\*Corresponding author: anna.ono@ruhr-uni-bochum.de*

**We report on a passively mode-locked Tm, Ho:CLNGG laser operating at 2090 nm, pumped by a multimode fiber-coupled 793 nm laser diode, representing the first demonstration of an affordable diode-pumped sub-100-fs mode-locked laser in the 2.1-µm wavelength range. Due to the disordered nature of the gain material that exhibits a broad and smooth gain profile, pulses as short as 88 fs were achieved with an average output power of 120 mW at a repetition rate of 70.3 MHz. The stable self-starting mode-locked operation was obtained using a semiconductor saturable absorber mirror based on GaSb material system.**

In the last decades, ultrafast lasers directly emitting in the short-wave mid-infrared region (1.5 – 3 µm) have gained attention to enable various applications such as silicon processing [1], nonlinear wavelength conversion to 3-15 µm based on non-oxide nonlinear crystals [2], terahertz generation by two-color laser-field induced plasma filaments [3], and high-harmonic generation towards the soft-X ray region [4] among others. Among various technologies explored in this wavelength region, 2.1 µm lasers are particularly attractive because this wavelength region coincides with a high-atmospheric transmission window, facilitating power and energy scaling as well as applications. Three-level transitions close to this attractive wavelength can be achieved with Tm and Ho ions, making laser materials based on these ions a topic of intense exploration. Figure 1 shows the overview of pulse duration and average output power of mode-locked bulk lasers based on $Tm^{3+}$- and/or $Ho^{3+}$-doped materials. Comparing Tm and Ho lasers, Ho lasers are particularly advantageous for high-power laser oscillators and amplifiers [5] [Fig.1 (purple)]. This is due to their high cross sections, long upper-level lifetimes, and small quantum defects due to the close pump (1.9 µm) and laser (2.1 µm) wavelengths. However, most Ho materials typically exhibit structured and narrow emission spectra making it difficult to reach sub-100-fs pulses in mode-locked operation [6]. In addition, one of the main drawbacks of Ho is the need for Tm laser pumping, not allowing for high-power laser diode (LD) pumping. With regards to Tm doped materials, they allow for high-power 790 nm AlGaAs LD pumping with a 2-for-1 pumping scheme achieved by cross-relaxation process. While they exhibit small cross sections and short lifetimes, their broad gain bandwidth enables few optical-cycle pulse generation [7,8] as shown in Fig. 1 (red). However, Tm typically has a gain bandwidth mostly on the shorter wavelengths (1.9-2.0 µm) compared to Ho (2.0-2.1 µm), which can make it susceptible to strong water-vapor absorption that exists below 2000 nm, except for sesquioxides [9].

In recent years, $Tm^{3+}$ and $Ho^{3+}$ co-doped materials, particularly those based on disordered host materials, have attracted significant attention for ultrashort pulse generation. Co-doping of Tm and Ho ions results in a red-shift of emission wavelength which can avoid a strong water-vapor-absorption band. Also, their slightly different emission wavelengths partially allow for broadening the gain. Furthermore, employing disordered host materials such as calcium aluminates [10], mixed sesquioxides [9,11], and multicomponent garnets [12,13], makes the gain profile broad and smooth via inhomogeneous spectral broadening, which enables the generation of few-cycle pulses. In common $Tm^{3+}/Ho^{3+}$ co-doped laser systems, the gain medium is pumped by using the $Tm^{3+}$ absorption band, and $Ho^{3+}$ ions are indirectly pumped by the energy transfer from $Tm^{3+}$ ions [14]. Therefore, this allows us to use readily accessible high-power 790 nm AlGaAs LDs as pump sources. Despite Tm or Tm,Ho co-doped lasers have the advantage of the availability of LD pumping, sub-100-fs mode-locked lasers have only been demonstrated using Ti:sapphire laser or 1.6 µm fiber lasers with excellent beam quality as pump sources so far [Fig. 1(red, green)]. The use of good beam quality pump lasers is beneficial to achieve soft-aperture Kerr-lens mode-locking which is the most promising for shortest pulse generation, as well as Kerr-lens assisted mode-locking which delivers shorter pulse durations overcoming the limitation of modulation depth of real saturable absorbers [15]. Particularly remarkable results were achieved by broad and smooth gain profiles and the great contribution of Kerr-lensing effect [7,8,16–18]. Although these are notable results, Ti:Sapphire lasers and 1.6 m fiber lasers are expensive and cumbersome, and ultrashort pulse lasers with LD pumping system is desirable for commercial products or industrial use. However, the performance of LD-pumped mode-locked lasers rather depends on the properties of the saturable absorber itself. So far, LD-pumped mode-locked Tm and Tm, Ho-based lasers were only demonstrated with longer pulse durations [19–21] [Fig.1 (blue)], and the shortest pulse duration of 170 fs was achieved by SESAM mode-locked Tm-singly doped $LuScO_3$ mixed sesquioxide single crystal laser [22].

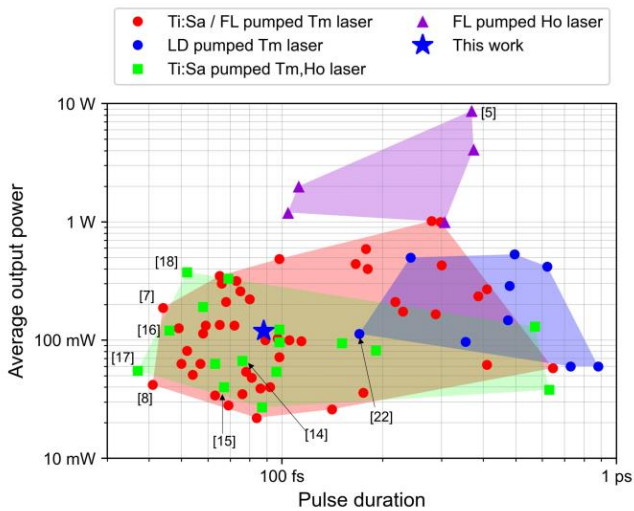

Fig. 1. Overview of femtosecond mode-locked bulk laser oscillators based on Tm and/or Ho doped materials. Each plot is categorized into Ti:sapphire (Ti:Sa) laser-, fiber laser (FL)-pumping, and LD-pumping.

In this study, we demonstrated a multimode LD-pumped passively mode-locked sub-100-fs Tm, Ho:CLNGG laser at 2.1-μm wavelength range for the first time. Due to the broad and smooth gain profile of the disordered gain material and a well-engineered GaSb-based SESAM, pulses as short as 88 fs with an average output power of 120 mW were achieved at a repetition rate of 70.3 MHz.

The experimental setup of the mode-locked Tm, Ho:CLNGG laser is depicted in Fig. 2(a). As a gain medium, we used a Tm,Ho co-doped CLNGG crystal. CLNGG is a family of calcium niobium gallium garnet (CNGG)-type crystal that exhibits a disordered crystal structure due to the randomly distributed $Nb^{5+}$ and $Ga^{3+}$ ions. That causes inhomogeneous broadening of absorption and emission spectral lines of dopant trivalent ions [12]. In CLNGG, the cationic vacancy concentration in CNGG is reduced by further doping univalent $Li^{1+}$ ions, improving the optical quality of the crystal [23]. For laser experiments, an anti-reflection (AR) coated $Tm^{3+}$ (2.34 at.%), $Ho^{3+}$ (0.54 at.%) co-doped CLNGG crystal with dimensions of 3×3×8 $mm^3$ was prepared. The crystal was wrapped with indium foil and mounted on a water-cooled copper holder. As a pump source, we used a commercially available multimode fiber-coupled LD emitting at 793 nm with a maximum output power of 5.4 W. Its core diameter and NA are 106.5 μm and 0.15, respectively.

The pump beam was collimated and focused using a doublet lens (f= 60 mm) and a plane-concave lens (f= 75 mm), resulting in a focused spot diameter of approximately 133 μm. The single-pass pump absorption of the laser crystal under non-lasing conditions was 74%.

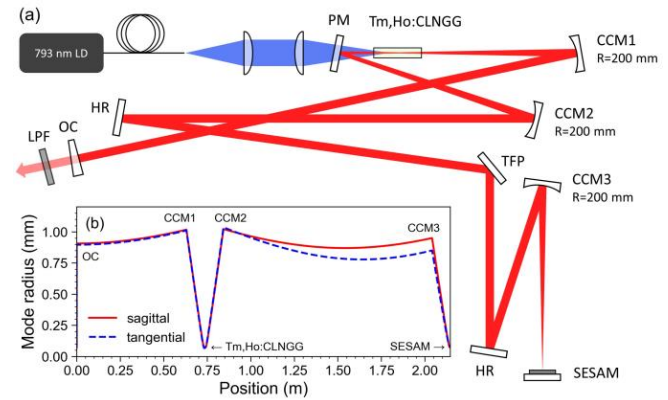

Fig. 2. (a) Schematic of the SESAM mode-locked Tm, Ho:CLNGG laser (PM: pump mirror, CCM: concave mirror, HR: high reflectivity mirror, TFP: thin film polarizer, OC: output coupler, LPF: long-pass filter to block residual pump beam), and (b) the calculated laser mode radius.

To obtain linearly polarized output, a thin-film polarizer (TFP) was used in the cavity since the pump source is unpolarized and the gain medium has a cubic structure. For mode-locked laser operation, we used a GaSb-based SESAM with a saturation fluence $F_{sat}$ of 8.58 μJ/$cm^2$ and a modulation depth $\Delta R$ of 0.93%, measured using an OPO with a pulse duration of 110-fs at a wavelength of 2050 nm [24,25], grown at ETH Zürich (see acknowledgments). The laser beam was focused on the SESAM using a concave mirror (R=200 mm). The round-trip group delay dispersion (GDD) of the cavity for 2.1 μm amounted to ~-850 $fs^2$ resulting from ~-300 $fs^2$ by cavity mirrors and the material dispersion of the gain crystal (-34 $fs^2$/mm at 2080 nm) [26]. The laser mode radius inside the cavity calculated by ABCD matrices is shown in Fig. 2(b). The laser mode radius on the laser crystal and the SESAM were 73×75 μm and 72×80 μm, respectively. The output beam passed through a long-pass filter to eliminate the residual pump beam.

Using an output coupler (OC) with a transmission of 1%, mode-locked laser experiments were performed. We confirmed the turnkey self-start operation with appropriate alignment. At a maximum incident pump power of 5.4 W, an average output power of 120 mW in mode-locked operation is obtained. The fluence on the SESAM under working condition was 943 μJ/$cm^2$. The measured optical spectrum is shown in Fig. 3(a). We observed a $sech^2$-shaped spectrum indicating soliton mode-locking at a center wavelength of 2090.6 nm with a spectral bandwidth (full width at half maximum) of 53.7 nm. This laser wavelength almost exceeds the longest emission wavelength edge of Tm:CLNGG ~2050 nm [27]. This is because in our case the Tm emission band is strongly affected by the Ho absorption band at the low inversion level because of the small cavity loss. Therefore, $Ho^{3+}$ gain contributes significantly to the laser emission, which explains the long center wavelength of 2090.6 nm. Moreover, this is related to the low optical-to-optical conversion efficiency of ~3%. Since all $Ho^{3+}$ ions are indirectly pumped by energy transfer from $Tm^{3+}$ ions, the effective inversion ratio is determined by the balance of the doping concentrations of both ions [28]. However, the doping concentrations of our crystal have not yet been optimized for efficient energy transfer. Therefore, the contribution of $Ho^{3+}$ ions, i.e., the gain at our laser wavelength is small. That led to low laser efficiency, however, this problem would be solved by optimizing the doping concentrations of both ions to obtain a more efficient energy transfer. The pulse duration was measured using an SHG-based autocorrelator (PulseCheck, A.P.E. GmbH), and the autocorrelation trace is presented in Fig. 3(b). Assuming $sech^2$-shaped pulses, the pulse duration was determined to be 88 fs. The time-bandwidth product was calculated to be 0.324, only slightly above the theoretical value of Fourier transform-limited $sech^2$ pulses. This residual chirp is in good agreement to that caused by material dispersion of the OC substrate (~-720 $fs^2$). The pulse energy and peak power calculated from a repetition rate of 70.3 MHz [Fig. 5(a)] reached 1.7 nJ and 19.4 kW, respectively.

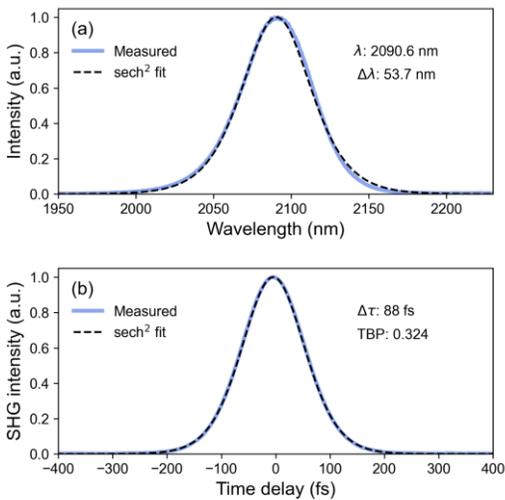

Fig.3. (a) Optical spectrum and (b) autocorrelation trace of the SESAM mode-locked Tm, Ho:CLNGG laser.

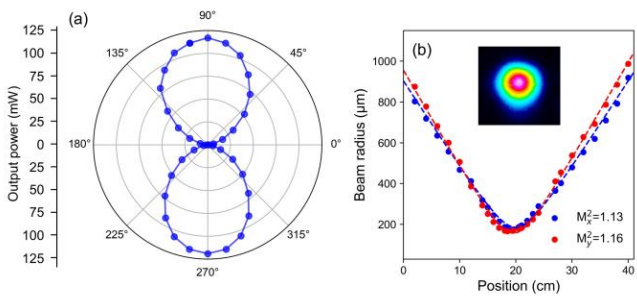

Fig. 4. (a) Average output power versus rotation of the polarizer, (b) $M^2$ measurement and the far-field beam profile (inset) of the SESAM mode-locked Tm, Ho:CLNGG laser.

We investigated the polarization characteristics of the mode-locked laser using a Glan-Taylor polarizer. Figure 4(a) shows the output power versus the rotation of the polarizer. A linearly polarized laser output was obtained owing to the TFP inside the cavity. The polarization extinction ratio was calculated to be 19.0 dB. In our experiments, we intentionally selected p-polarization using the TFP because the laser output without the polarization selective element exhibited slight vertical polarization, which can be attributed to stress in the gain crystal. Figure 4(b) shows the beam quality $M^2$ measurement according to ISO 11146 standard with an IR camera (WinCamD, DataRay). From the fitting curves, the measured $M^2$ values were 1.13 and 1.16 in sagittal and tangential planes, respectively. The far-field beam profile [Fig. 4(b) inset] presented a close to $TEM_{00}$ Gaussian mode profile.

Steady-state single-pulse mode-locking was confirmed by radio frequency (RF) spectra with different span ranges. The spectra were measured by an RF spectrum analyzer (FPC1000, Rohde & Schwarz). Figure 5(a) shows the fundamental beat note at 70.29 MHz, with a signal-to-noise ratio of 58.5 dB. The mode-locked pulses did not show any Q-switching instability. Figure 5(b) presents harmonic beat notes for a wide span. The beat notes exhibit almost the same amplitude for a 1-GHz spanning range. Moreover, we measured a waveform of mode-locked pulses [Fig. 5(c)] with a 12.5-GHz photodiode (ET-5000, Coherent Inc.) recorded with a 25-GHz sampling oscilloscope (PicoScope 9000, Pico Tech.). No pulses were observed between the two pulses with a temporal separation of ~14 ns, corresponding to the round-trip time of the cavity. In addition, 16-ps span autocorrelation trace in Fig. 5(d) showed the absence of pre- or post-pulses. These results indicate that single-pulse operation was obtained.

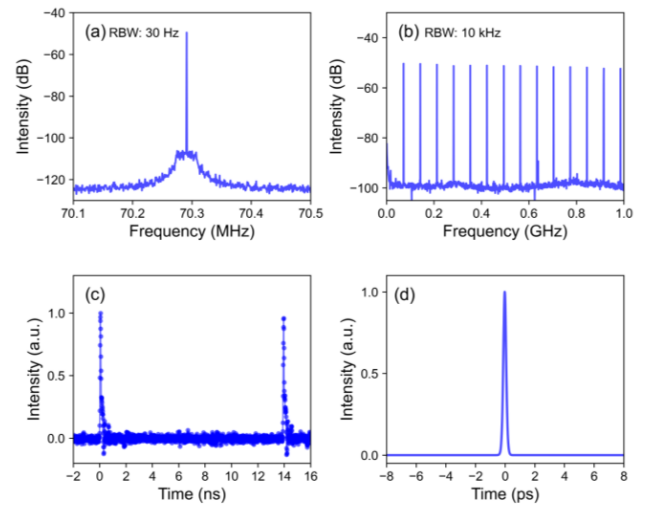

Fig. 5. Radio frequency spectra (a) at the fundamental beat note and (b) in a span of 1 GHz recorded with the resolution bandwidth (RBW) of 30 Hz and 10 kHz, respectively. (c) Waveform measured with a 12.5-GHz photodiode and a 25-GHz sampling oscilloscope. (d) Autocorrelation trace in a span of 16 ps.

In conclusion, we achieved a sub-100-fs multimode LD-pumped passively mode-locked laser at 2.1 μm wavelength for the first time. The laser was self-starting and passively mode-locked using a GaSb-based SESAM. Pulses as short as 88 fs were achieved with an average output power of 120 mW at the repetition rate of 70.3 MHz. The broad and smooth gain profile of the Tm,Ho:CLNGG crystal supported this short pulse generation. The maximum output power is limited by the available pumping power and the laser efficiency which is determined by the efficiency of energy transfer in the pumping scheme. Alternatively, using a higher transmission output coupler and increasing cavity GDD can be a quick solution for power scaling, however, a longer pulse duration would be expected in that case. In the multimode LD-pumped laser system, the laser efficiency basically falls behind that of a single-mode laser pumping owing to worse mode-matching efficiency. Nevertheless, this disadvantage can be covered by high output powers and low cost of commercially available LDs emitting up to hundreds of watts. Our results demonstrate a novel 2.1 µm ultrafast laser based on an affordable LD-pumping system, which is compatible with the conventional Ti:Sapphire laser or fiber laser pumped laser systems, that paves the way for LD-pumped high power laser oscillators in a highly coveted spectral region.

**Funding.** These results are part of a project that has received funding from the European Research Council (ERC) under the European Union's Horizon 2020 research and innovation programme (grant agreement No. 805202 - Project Teraqua), and National Natural Science Foundation of China (52072351), Qilu Young Scholar Program of Shandong University, Taishan Scholar Foundation of Shandong Province.

**Acknowledgments.** We would like to thank Shahwar Ahmed and Mustafa Hamdan for their early support in setting up the setup. We acknowledge the Ultrafast Laser Physics Group at ETH Zürich for providing and characterizing the SESAM employed in this study.

**Disclosures**. The authors declare no conflicts of interest.

**Data availability.** Data underlying the results presented in this paper are not publicly available at this time but may be obtained from the authors upon reasonable request.